\documentclass[aps,prx,twocolumn,preprintnumbers,superscriptaddress,10pt,showkeys]{revtex4-2}
\usepackage{amsmath}
\usepackage{amssymb}
\usepackage{float}
\usepackage{hyperref}
\usepackage{nicefrac}
\usepackage[dvipsnames]{xcolor}
\usepackage{cancel}
\usepackage[capitalise]{cleveref}
\usepackage{booktabs}
\usepackage{mathtools}

\newcommand{\dd}{\mathrm{d}}

\makeatletter
\g@addto@macro\bfseries{\boldmath}
\makeatother

\usepackage{graphicx}

\begin{document}

\title{Strong model-agnostic constraints for twin-star solutions}

\author{Sofia Blomqvist}
\email{sofia.blomqvist@helsinki.fi}
\affiliation{Department of Physics and Helsinki Institute of Physics, P.O.~Box 64, FI-00014 University of Helsinki, Finland}
\author{Christian Ecker}
\email{ecker@itp.uni-frankfurt.de}
\affiliation{Institut f\"ur Theoretische Physik, Goethe Universit\"at, Max-von-Laue-Str. 1, 60438 Frankfurt am Main, Germany}
\author{Tyler Gorda}
\email{gorda.1@osu.edu}
\affiliation{Center for Cosmology and AstroParticle Physics (CCAPP), Ohio State University, Columbus, OH 43210}
\affiliation{Department of Physics, The Ohio State University, Columbus, OH 43210, USA}
\author{Aleksi Vuorinen}
\email{aleksi.vuorinen@helsinki.fi}
\affiliation{Department of Physics and Helsinki Institute of Physics, P.O.~Box 64, FI-00014 University of Helsinki, Finland}
\preprint{HIP-2025-26/TH}

\begin{abstract}
We perform a model-agnostic Bayesian analysis of the neutron-star-matter equation of state (EoS), using known ab-initio constraints and astrophysical observations to limit its behavior at intermediate densities.
Permitting explicit first-order phase transitions allows us to systematically search for twin-star solutions, i.e.~the existence of stars degenerate in mass but differing in radius.
We find that current observational constraints exclude all but two classes of twin stars. 
The first is characterized by a first-order transition occurring at a very low density, where the material properties of the system either stay largely intact or move away from the conformal limit.
In the second, more interesting class, the discontinuity in the mass-radius curve emerges after a rapid crossover transition at a significantly higher density, with the speed of sound exhibiting two sharp peaks at distinct densities. 
Since neither class shows clear conformalization upon entering the second branch, the standard twin-star scenario linking the mass–radius discontinuity to deconfinement can be firmly ruled out, while even the remaining solutions —-- disfavored by per-mille Bayes factors and in tension with theoretical bounds --— are likely to be excluded in the future.
\end{abstract}

\maketitle

\textit{Introduction.}---Advances in theoretical ab-initio studies of nuclear \cite{Tews:2012fj,Lynn:2015jua,Drischler:2017wtt,Drischler:2020hwi,Drischler:2020yad,Keller:2022crb,Alp:2025wjn} and quark matter \cite{Kurkela:2009gj,Gorda:2021kme,Gorda:2021znl,Gorda:2023mkk,Karkkainen:2025nkz} as well as neutron-star (NS) observations \cite{Demorest:2010bx,Antoniadis:2013pzd,Fonseca:2016tux,Cromartie:2019kug,Fonseca:2021wxt,TheLIGOScientific:2017qsa,LIGOScientific:2018cki,Shawn:2018,Steiner:2017vmg,Nattila:2017wtj, Miller:2019cac,Riley:2019yda,Miller:2021qha,Riley:2021pdl} have recently enabled a model-agnostic approach to inferring the thermodynamic properties of NS matter, including its equation of state (EoS). 
This has led to crucial insights of both quantitative and qualitative nature, including a robust demonstration of the rapid stiffening of the EoS at low and moderate densities, corresponding to a high maximum speed of sound \cite{Hebeler:2013nza,Kurkela:2014vha,Annala:2017llu,Most:2018hfd,Tews:2018iwm,Landry:2018prl,Capano:2019eae,Miller:2019nzo,Essick:2019ldf,Raaijmakers:2019dks,Dietrich:2020efo,Landry:2020vaw,AlMamun:2020vzu,Essick:2021kjb,Raaijmakers:2021uju,Annala:2021gom,Huth:2021bsp,Altiparmak:2022bke,Lim:2022fap,Gorda:2022jvk}. At even higher densities, both hard-cutoff and Bayesian studies have identified a marked softening of the EoS, leading to near-conformal behavior below the central densities of maximal-mass NSs.
This has been interpreted as signaling a transition to deconfined degrees of freedom within Quantum Chromodynamics (QCD) \cite{Annala:2019puf,Annala:2023cwx,Komoltsev:2023zor,Ayriyan:2025rub}.

Despite the progress described above, some important caveats remain. 
A prominent example has to do with the possibility of a discontinuous first-order phase transition (FOPT) between two phases of high-density QCD matter --- be that two separate phases of confined matter or, more interestingly, hadronic and quark matter (QM).
Strong FOPTs, featuring a sizable jump in energy density, are known to have potentially interesting phenomenological consequences both for binary NS mergers~\cite{Bauswein:2018bma,Most:2018eaw,Ecker:2019xrw,Weih:2019xvw,Most:2019onn,Tootle:2022pvd,Fujimoto:2022xhv,Ecker:2024kzs} and quiescent NSs \cite{Chamel:2012ea,Alford:2013aca,Demircik:2020jkc,Brandes:2023hma,Brandes:2023bob,Ecker:2025vnb}. 
Only a handful of works have, however, attempted to constrain FOPTs in a model-agnostic fashion \cite{Gorda:2022lsk,MUSES:2023hyz,Essick:2023fso,Brandes:2023hma,Mroczek:2023zxo,Komoltsev:2024lcr,Verma:2025dez,Tang:2025xib,Li:2025obt}, and even fewer such studies have systematically studied a specific phenomenon often linked to FOPTs, i.e., the existence of two disjoint branches of compact stars (see, however, \cite{Montana:2018bkb} for an early attempt in this direction). 
Intriguingly, this phenomenon gives rise to stellar solutions sharing a common mass but differing in radii, commonly referred to as twin stars \cite{Gerlach:1968zz}, which offer a clear observational prediction for current and future mass-radius measurements.

In more phenomenological studies, twin stars have received extensive attention both in the astrophysics and nuclear-physics communities (see e.g.~\cite{SchaffnerBielich2002,Alford2005,Benic:2014jia,Alvarez-Castillo:2016oln,Alvarez-Castillo:2017qki,Li:2024sft} and references therein), leading to the identification of multiple classes of potential solutions \cite{Montana:2018bkb}. 
Given recent advances in both the observational study of NSs and ab-initio limits from nuclear and particle theory, it is, however, unclear whether these results are still valid and if all proposed classes of twin stars continue to be viable.

This situation motivates us to address the viability of twin stars in a maximally robust and conservative manner, only relying on commonly accepted observational data and first-principles EoS limits.
To this end, we introduce an EoS ansatz versatile enough to describe all twin classes previously identified and perform a Bayesian analysis involving a gradual introduction of astrophysical constraints. 
As detailed in our letter, this leads us to a crisp conclusion: current observations rule out all but two of the twin categories suggested in \cite{Montana:2018bkb}, with the viable EoSs featuring either an early-onset FOPT beginning around $1.5$ nuclear saturation densities $n_{\rm s} \approx 0.16~\mathrm{fm}^{-3}$ or a crossover-type transition at 4-5~$n_{\rm s}$. 
Given that  the Bayes factors for both solutions are at the per-mille level, and that neither family shows signs of conformalization upon entering the second branch, we conclude that deconfinement-induced twin solutions can be firmly ruled out.
Finally, we analyze how various observational constraints affect our conclusion, finding X-ray data to play a particularly prominent role in the process.

\textit{Setup.}---We parameterize our EoSs in three parts.
For baryon-number densities $n < 0.5~n_{\rm s}$, we apply the Baym-Pethick–Sutherland (BPS) prescription for the NS crust~\citep{Baym:1971pw}, while in the range $0.5\,n_{\rm s} \le n < n_{\rm CEFT}$, we sample polytropic EoSs with pressure values $p_{\rm CEFT} \in [1.5, 7]\,\mathrm{MeV}/\mathrm{fm}^3$, uniformly distributed at $n_{\rm CEFT} = 1.1\,n_{\rm s}$. 
This covers the conservative next-to-next-to-leading order (N2LO) Chiral Effective Field Theory (CEFT) uncertainty band from~\cite{Drischler:2020fvz}.

The third component of the EoS is built from squared-speed-of-sound functions $c
_s^2(\mu)$, taken to be piecewise linear in the baryon chemical potential $\mu$~\cite{Annala:2019puf},
\begin{equation*}
    c_{\rm s}^2(\mu)=\frac{\left(\mu _{i+1}-\mu \right) c_{{\rm s},i}^2+\left(\mu -\mu_i\right) c_{{\rm s},i+1}^2{}}{\mu _{i+1}-\mu _i}\,,\ \mu_i\leq\mu\leq\mu_{i+1}\,,
\end{equation*}
with the parameters $c^2_{{\rm s,}i}\in[0,1]$ and $\mu_i\in[\mu_{\rm CEFT},\mu_{\rm pQCD}]$  uniformly sampled.
The functions thus built are required to be consistent with the perturbative-QCD (pQCD) results of~\cite{Gorda:2021znl,Gorda:2021kme,Gorda:2023mkk} at $\mu_{\rm pQCD} = 2.6\,\text{GeV}$. 
This is implemented through matching in both $n$ and $p$, related to the speed of sound through the thermodynamic relations
\begin{align}
    n(\mu)&=n_{\rm CEFT}\,\exp \left[\int_{\mu_{\rm CEFT}}^\mu \frac{\dd\mu'}{\mu'c_\mathrm{s}^2(\mu')} \right]\,,\label{nint}\\
    p(\mu)&= p_{\rm CEFT}+\int_{\mu_{\rm CEFT}}^\mu \dd\mu' n(\mu')\,. \label{eq:np}
\end{align}

An important detail in our implementation of the pQCD constraint is that we perform a scale averaging introduced in \cite{Gorda:2023usm} by sampling the modified minimal-subtraction renormalization scale $\bar{\Lambda} \in [1/2, 2] \times (2 \mu / 3)$ with a log-uniform weight.
We also fold in information about the well-converging speed of sound at lower densities by sampling two additional points $c^2_{{\rm s},i}$ at $\mu_i = 2.25, 2.43\,\mathrm{GeV}$  \cite{Komoltsev:2023zor}. 
The values used here are chosen from normal distributions with mean and standard deviation given by the scale-averaged mean and twice the standard deviation of the corresponding perturbative $c_{\rm s}^2(\mu_i)$.

The EoS ansätze constructed in this way depend on a total of $N_p = 9$ parameters, of which two are used to match to the CEFT and pQCD limits, while the remaining seven parameterize either six intermediate segments in $c_\mathrm{s}^2$ or five segments and one FOPT. 
We implement the FOPT as a discontinuity at equal chemical potentials, $\mu_k = \mu_{k+1} \eqcolon \mu_{\rm PT}$, with $\mu_{\rm PT}$ sampled uniformly in $[\mu_{\rm CEFT}, \mu_{\rm pQCD}]$.
Between these points, we introduce a (uniform) random jump in the number density with $\Delta n/n_{\rm s} \in [0.1, 12]$.
The (baryon) number density $n^+(\mu_{\rm PT})=n^-(\mu_{\rm PT})+\Delta n$ and energy density $e^+(\mu_{\rm PT})=e^-(\mu_{\rm PT})+\mu_{\rm PT}\Delta n$ are then adjusted accordingly while the pressure is kept constant, $p^+(\mu_{\rm PT})=p^-(\mu_{\rm PT})$.

To assess the likelihood of each EoS, we use Bayes' theorem to relate it to the product of uncorrelated likelihoods derived from CEFT and astrophysical observations. This amounts to the relation
\begin{align}\label{eq:posterior}
P(\text{EoS}|\text{data}) \propto{}& P(\text{CEFT}|\text{EoS}) P(\text{Mass}|\text{EoS}) \nonumber\\
&\times P(\tilde{\Lambda}|\text{EoS}) P(\text{X-ray}|\text{EoS})\,,
\end{align}
where $P(\text{CEFT}|\text{EoS})$ denotes the Gaussian likelihood inferred from the N2LO calculation of~\cite{Drischler:2020fvz} and the latter three terms correspond to mass, tidal-deformability, and X-ray radius measurements, reviewed in Appendix \ref{sec:obs}.

For a given EoS with multiple stable branches, the likelihood $P(\text{data} | \text{EoS})$ is computed as a sum of likelihoods for each stable branch, or in the case of the GW constraint, a sum over all pairs of branches.
The likelihood for each branch is evaluated following a procedure outlined in \cite{Annala:2023cwx} (see also \cite{Gorda:2022jvk}), taking the prior for the mass to be flat on the interval $[\max(0.5M_\odot, M_\text{min}), M_\text{max}]$, where $M_\text{min}$ and $M_\text{max}$ are the minimum and maximum masses on a given EoS branch.

\begin{figure*}[t!]
    \centering
    \includegraphics[width=\linewidth]{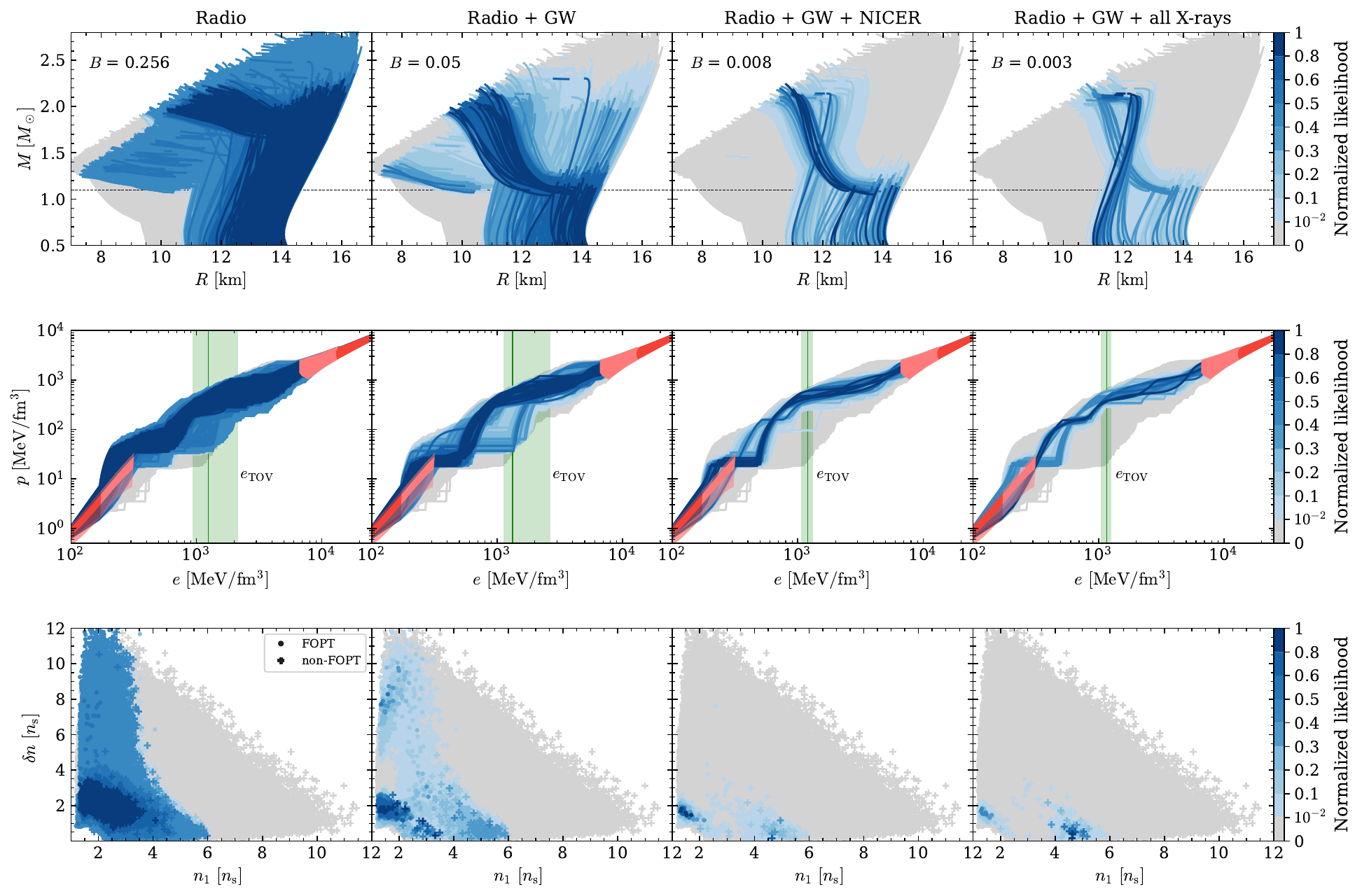}
    \caption{
    Progressive impact of astrophysical constraints (from left to right) on the mass-radius relation (top), the EoS (middle), and the correlation between the parameters $n_1$ and $\delta n$ (bottom) for twin-star solutions.
    Shading from light to dark blue indicates increasing normalized likelihood (calibrated separately for each column), with values smaller than $10^{-2}$ shown in gray, while the FOPT- and non-FOPT-induced twin-star solutions are distinguished by circular and cross markers on the third row.
    Green lines and bands in the EoS panels show the median and $68\%$ credible interval of the central density in maximally massive stars.
    At high densities, the dark and light red regions represent the pQCD constraints imposed on the pressure and sound speed, respectively, while at low densities, the dark and light red bands correspond to CEFT pressure constraints up to $1.1\,n_s$ (imposed on our results) and $2\,n_s$ (shown here for illustrative purposes), respectively. For the latter results, the narrow non-transparent bands correspond to 1-$\sigma$ and the wider transparent ones to 2-$\sigma$ estimates.
    }
    \label{fig:constraints}
\end{figure*}

To quantify how well twin-star solutions are supported by astrophysical data, we determine the Bayes factor between twin and non-twin EoS classes (referred to as models below), $B \coloneq \frac{P(\text{data} | \text{twin})}{P(\text{data} | \text{non-twin})}$, where each model likelihood is obtained by marginalizing over all model parameters.
In this context, a twin-star model is defined as an EoS that produces a mass-radius sequence with two distinct stable branches, allowing for at least some static equilibrium solutions of the Tolman-Oppenheimer-Volkoff (TOV) equations that share the same mass ($M_1 = M_2$) but differ in radii ($R_1 \neq R_2$). 
Here, we assume a conservative lower mass bound $1.1~M_\odot$ for the maximum mass of the first branch, and have verified that imposing an even more conservative bound of $1~M_\odot$ has no qualitative effect on our results or conclusions.
This is done to ensure consistency of both branches with the lower mass limit inferred from neutron-star formation analyses~\cite{Muller:2024aod} ($M = 1.192~M_\odot$), as well as with the mass of the lightest known neutron star with a precise measurement --- the $M = 1.174~M_\odot$ companion in the eccentric compact binary system PSR J0453+1559~\cite{Martinez:2015mya}. 
As usual, $B\gg 1$ would indicate strong support for the twin hypothesis (i.e.~that the NS-matter EoS realized in nature supports two disjoint branches), $B\approx 1$ comparable support for both hypotheses, and $B\ll 1$ support for the standard scenario.
All necessary integrals are evaluated numerically by Monte-Carlo averaging over our discrete EoS samples, assigning each sampled EoS an equal prior weight $1/N_\text{EoS}$, where $N_{\text{EoS}}$ is the number of EoSs in the model class.

Finally, as noted in, e.g., \cite{Zhou:2025uim,Papadopoulos:2025uig}, twin-star solutions can arise not only from genuine FOPTs but also from smooth EoSs that exhibit features resembling FOPTs, such as regions where the sound speed drops sharply to small, though nonzero values.
For this reason, we will classify our results not in terms of the phase-transition parameters, such as the onset density  $n_{\rm PT}:=n^-(\mu_{\rm PT})$ or the discontinuity $\Delta n$, but through the density $n_1$ reached when the first branch becomes unstable and the distance to the start of the second branch, $\delta n$.

\begin{figure*}[t]
    \centering
    \includegraphics[width=\linewidth]{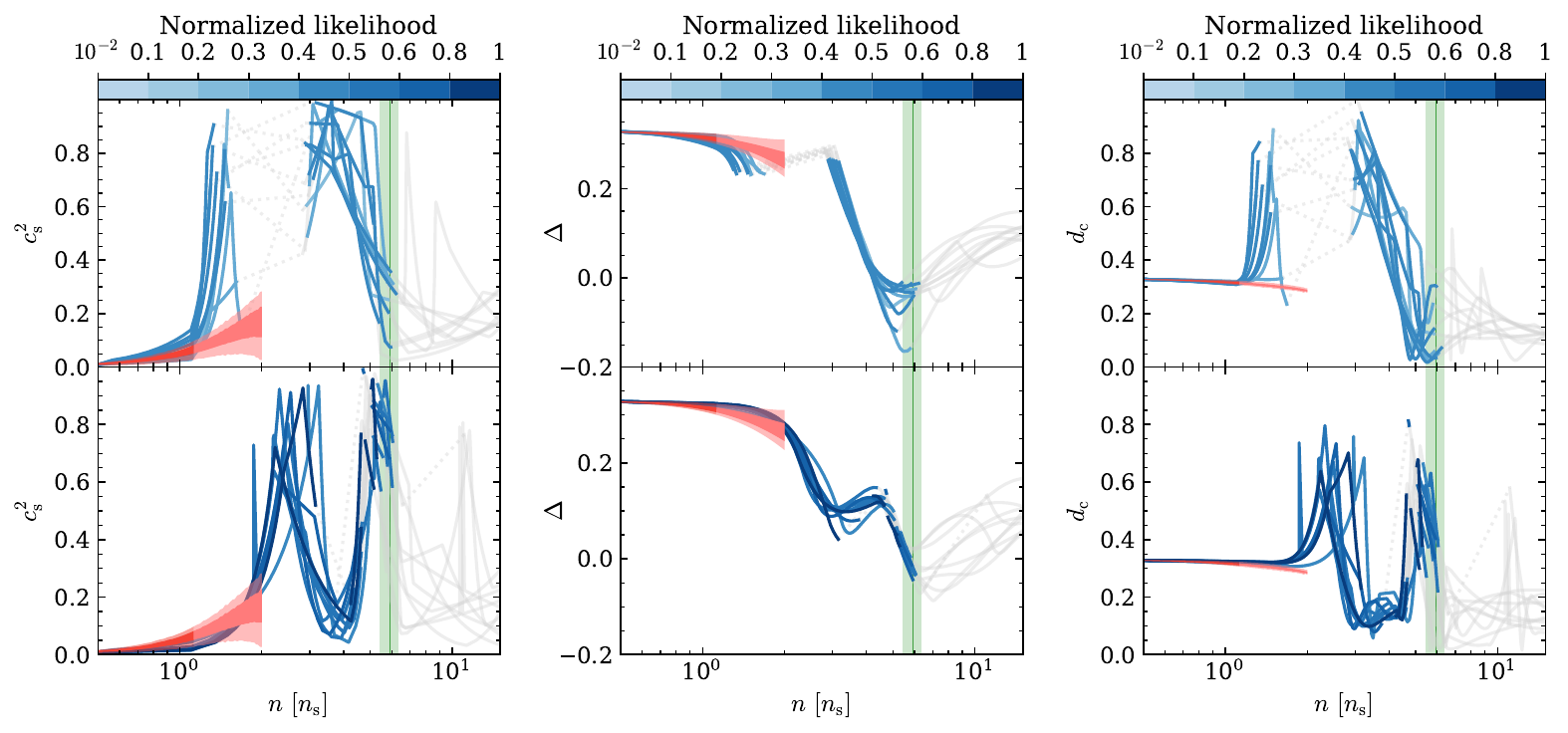}
    \caption{Ten highest-likelihood twin-star EoSs from twin categories 1 (upper row) and 2 (lower row). From left to right we show the squared sound speed ($c_{\rm s}^2$), conformal anomaly ($\Delta$) and conformal distance ($d_{\rm c}$). Green lines and bands show the median and $68\%$ credible interval of the central density of maximally massive stars, while the N2LO CEFT error bands are shown as in Fig.~\ref{fig:constraints}~\cite{Drischler:2020fvz}.
    Finally, we note that the blue line segments correspond to stable twin-star branches, solid gray lines to unstable parts, and dotted lines to the phase transition region.
     }
    \label{fig:EoSproperties}
\end{figure*}

\textit{Results.}---Figure~\ref{fig:constraints} provides a summary of our main results. 
The three rows display the mass-radius relations, EoS, and $n_1$--$\delta n$ distributions for our ensemble of approximately $10^7$ models, of which about $2.7\times10^5$ exhibit twin-star solutions.
The corresponding Bayes factors are indicated in the upper-left corners of the top-row panels.
From the first to the fourth column, we progressively incorporate astrophysical constraints, beginning with radio measurements of massive pulsars, then adding tidal-deformability data from the GW170817, X-ray radius measurements from NICER, and finally all available X-ray data summarized in Table~\ref{table:measurements}. 
Throughout, we apply the CEFT and pQCD constraints and indicate the relative likelihoods of the solutions with a color coding, where the highest-likelihood EoSs are shown in darkest blue. 
The gray background visible in each panel displays our prior distribution, dictated by the low- and high-density limits of the EoS, causality, and thermodynamic consistency~\footnote{The nearly linear upper edge of the $n_1$-$\delta n$ prior, corresponding to a slope of $-1$, arises from the upper bound of our prior on $\Delta n \leq 12\,n_s$. Since our posterior distributions are peaked far from this edge, this cut does not impact our results.}.

The first key insight from these results is related to the trend of the Bayes factor: while already low when only radio measurement are taken into account, each new set of observational data progressively reduces the number, indicating that twin-star EoSs are clearly disfavored by current astrophysical data.
In particular, the tidal-deformability data excludes many solutions that predict large radii for massive NSs along the fist branch, while the NICER radius measurements eliminate the small-radius low-mass twins identified in \cite{Gorda:2022lsk}.
While the final value $B = 0.003$ in the rightmost panel, obtained when all known theoretical and astrophysical constraints are taken into account, does not strictly exclude twin stars, it allocates only marginal evidence for their existence.

Concentrating next on the viable twin EoSs, shown in blue color in the rightmost panel, we note that they form two distinct classes (see also Appendix \ref{sec:AppendixB}), in which:
\begin{enumerate}   
\item The first branch ends at a FOPT \footnote{Note that there is a small number of class-1 EoSs, virtually indistinguishable from the rest, where the transition is a very rapid crossover instead of a genuine FOPT.} occurring at a low density and resulting in
\begin{equation}
    n_1 = 1.43^{+0.20}_{-0.11}~n_{\rm s}\,,\quad \delta n = 1.55^{+0.29}_{-0.30}~n_{\rm s}\,,
\end{equation}
where all values are quoted at $68\%$ credibility.
For these EoSs, the first branch barely exceeds our lower bound for the neutron-star mass $M\approx1.1~M_\odot$, while the twin branch starts from even lower masses, is centered around $R \approx 12~\text{km}$, and terminates near $M \approx 2.2~M_\odot$. 
\item The first branch terminates after a rapid crossover transition that significantly softens the EoS, taking place at considerably higher densities, with
\begin{equation}
n_1 = 4.64^{+0.37}_{-0.30}~n_{\rm s}\,,\quad \delta n = 0.55^{+0.37}_{-0.20}~n_{\rm s}\,.
\end{equation}
For these solutions, the first branch extends all the way to two solar masses, while the twin branch is nearly horizontal with a fixed mass around $2M_\odot$. 
It is worth noting that most EoSs in this class exhibit a FOPT at even higher densities, which is necessary to reach the pQCD limit in a causal fashion.
\end{enumerate}
In the EoS plot, the light red shading at high densities marks the region where the pQCD sound-speed constraint is imposed, while at low densities the light red extension up to $2~n_\mathrm{s}$ illustrates the CEFT error band~\cite{Drischler:2020fvz} that we do not not impose here. 
It is worth noting that both classes of viable twin-star EoSs are largely incompatible with the latter constraint, so it is conceivable that all remaining twin solutions will be eventually ruled out.

In Fig.~\ref{fig:EoSproperties}, we next inspect various features of the viable twin-star EoSs, displaying only ten highest-likelihood representatives of both classes for clarity.
The first panel shows the behavior of the squared sound speed as a function of baryon number density.
A notable feature in all viable twin-star EoSs is the steep rise of $c_{\rm s}^2$ at low densities, reaching peak values near the causal limit.
For the class-1 EoSs, the peak in the sound speed is followed by a FOPT, beyond which the EoS continues very stiff, with $c_{\rm s}^2(n_{\rm PT}+\Delta n) \gtrsim c_{\rm s}^2(n_{\rm PT})$.
At even higher densities, the sound speed drops sharply and remains low, $c_{\rm s}^2 \lesssim 0.2$, which is necessary for the quantity to approach the asymptotic conformal limit $c_{\rm s}^2=1/3$ from below in a thermodynamically consistent manner.
For the class-2 EoSs, we witness an interesting two-peak structure, where the speed of sound needs to first increase to ensure reaching the two-solar-mass limit but then soften and stiffen again to generate the crossover behavior that leads to the destabilization of the first branch and the onset of the second one. 
We note that both classes of EoSs are in visible tension with the CEFT bound for the sound speed at the lowest densities \cite{Drischler:2020fvz}, which will likely increase with the introduction of more accurate CEFT results in the future.

The second panel shows the corresponding behavior of the conformal anomaly $\Delta \coloneq 1/3 - p/e$~\cite{Fujimoto:2022xhv}, i.e., the normalized trace of the energy momentum tensor, which vanishes in conformal systems, such as deconfined quark matter at asymptotically high densities.
As shown in several studies~\cite{Fujimoto:2022xhv,Marczenko:2022jhl,Ecker:2025vnb}, the conformal anomaly of NS matter first monotonically decreases from its low-density value of $\Delta \approx 1/3$.
For our class-1 EoSs, this decrease is somewhat faster than the behavior predicted by CEFT \cite{Drischler:2020fvz}, reaching $\Delta(n_{\rm PT}) \approx 0.25$ at the onset of the FOPT and moving slightly upwards over the transition.
Considering that this value is significantly larger than the limiting value suggested for deconfined quark matter in previous model-agnostic analyses~\cite{Annala:2023cwx} and also contrasts with results from microphysical descriptions of a FOPT between nuclear and quark matter~\cite{Ecker:2025vnb}, we conclude that this transition would very likely occur between two distinct phases of hadronic matter.
For the class 2 EoSs, $\Delta$ on the other hand decreases more slowly, in better agreement with the CEFT results, with the value briefly saturating around 0.1 just prior to the destabilization of the first branch, and then continuing a decrease along the second branch.

As a final indicator, we show the behavior of the conformal distance $d_{\rm c} \coloneq \sqrt{\Delta^2 + (\Delta')^2}$, $\Delta' \coloneq \dd \Delta / \dd \log e$, in the last panel of Fig.~\ref{fig:EoSproperties}. 
This quantity was introduced in Ref.~\cite{Annala:2023cwx} as an improved metric for conformality that only vanishes when both $\Delta$ and its rate of change tend to zero.
A comprehensive comparison with model predictions further suggested the limit $d_{\rm c}\leq 0.2$ as a criterion for the presence of deconfined degrees of freedom, which has recently found support from a microphysical description of the transition~\cite{Ecker:2025vnb}. 
As expected, a closer inspection of this quantity verifies that the class-1 EoSs likely describe a transition between two hadronic phases, while the case of class-2 EoSs is more peculiar, as they exhibit a transition consistent with deconfinement, followed by a second transition moving away from conformal behavior.

\textit{Conclusion.}---In the letter at hand, we have constructed a large ensemble of parameterized EoSs for cold, beta-equilibrated neutron-star matter, roughly half of which  feature an explicit first-order phase transition. 
The EoSs are, by construction, consistent with theoretical bounds from perturbative QCD at high densities and are constrained by uncertainty estimates from chiral effective field theory at low densities.
The seven-segment parametrization, combined with a broad range of onset densities $n_{\rm PT}/n_{\rm s} \in [1.1, 40]$ and phase transition strengths $\Delta n/n_{\rm s} \in [0.1, 12]$, makes this large ensemble sufficiently general to faithfully represent all theoretically allowed NS EoSs with a FOPT.

In particular, we can distinguish between so-called twin-star models that produce mass-radius sequences featuring at least two stable stars with the same mass (above the expected formation threshold $M > 1.1\,M_\odot$) but different radii, and non–twin solutions that do not meet this criterion.
By progressively imposing observational constraints from the GW170817 event, precise mass measurements of heavy pulsars, and X-ray radius measurements by NICER and other collaborations, we show that the parameter space for viable twin-star solutions is drastically reduced.
We find the Bayes factor for twin-star models to be very small, $B=0.003$, indicating that they are strongly disfavored by astrophysical data.
The viable twin-star EoSs fall into two distinct classes, either featuring an early phase transition with an onset density $n_{\rm PT}\approx 1.5\,n_{\rm s}$ and a transition strength $\Delta n\approx 1.3\,n_{\rm s}$ or a crossover-type transition centered around 4~$n_{\rm s}$, both with little allowed variance.

The fact that the phase transition required for the first class of twin stars appears to occur between two distinct phases of nuclear matter is in strong tension with recent CEFT results~\cite{Drischler:2020fvz}, which provide error estimates for the EoS up to approximately $2\,n_{\rm s}$.
To this end, it is conceivable that the first class of twin-star solutions will be firmly ruled when more stringent low-density constraints are released in the future.

The material properties of the second class are also in tension with the CEFT results, a fact particularly pronounced in the $c_{\rm s}^2$ and $\Delta$  predictions near $n_{\rm s}$.
Moreover, while necessary for the generation of the second branch of compact stars, the two-peak structure in the squared speed of sound and the fluctuation of $\Delta$ and $d_{\rm c}$ between near-conformal and non-conformal behavior are at odds with general expectations for these quantities and may offer a way to rule out this class in the future, too.

The final take-away message from our results is twofold.
On one hand, the overall Bayes factor $B\approx 0.003$ for twin-star models demonstrates that their existence is strongly disfavored but not entirely ruled out by current astrophysical data.
At the same time, we have shown that the two classes of EoSs that account for this small but nonzero evidence have highly fine-tuned properties, predicting either a FOPT at a very low density or a distinctive two-peak structure for the speed of sound that can be either verified or ruled out.
Finally, we note that while the present analysis has focused specifically on twin-star solutions, the methods we have developed for this work are directly applicable to a generic analysis of first-order phase transitions in the NS-matter EoS, to which we shall return in the future.

\textit{Acknowledgments.}---We thank David Blaschke for repeatedly encouraging us to perform a model-agnostic study of twin-star solutions, as well as Oleg Komoltsev, Aleksi Kurkela, Joonas Nättilä, Luciano Rezzolla, and Andreas Schmitt for useful discussions. SB and AV acknowledge support from the Research Council of Finland, grant 354533, CE and TG from the Deutsche Forschungsgemeinschaft (DFG, German Research Foundation) through the CRC-TR 211 `Strong-interaction matter under extreme conditions'-- project number 315477589 -- TRR 211. TG also acknowledges support from the ERC AdG ``JETSET: Launching, propagation and emission of relativistic jets from binary mergers and across mass scales'' (Grant No.~884631).

\appendix

\section{Phase-transition and branch parameters}\label{sec:AppendixB}

\begin{figure}[t]
    \centering
    \includegraphics[width=85mm]{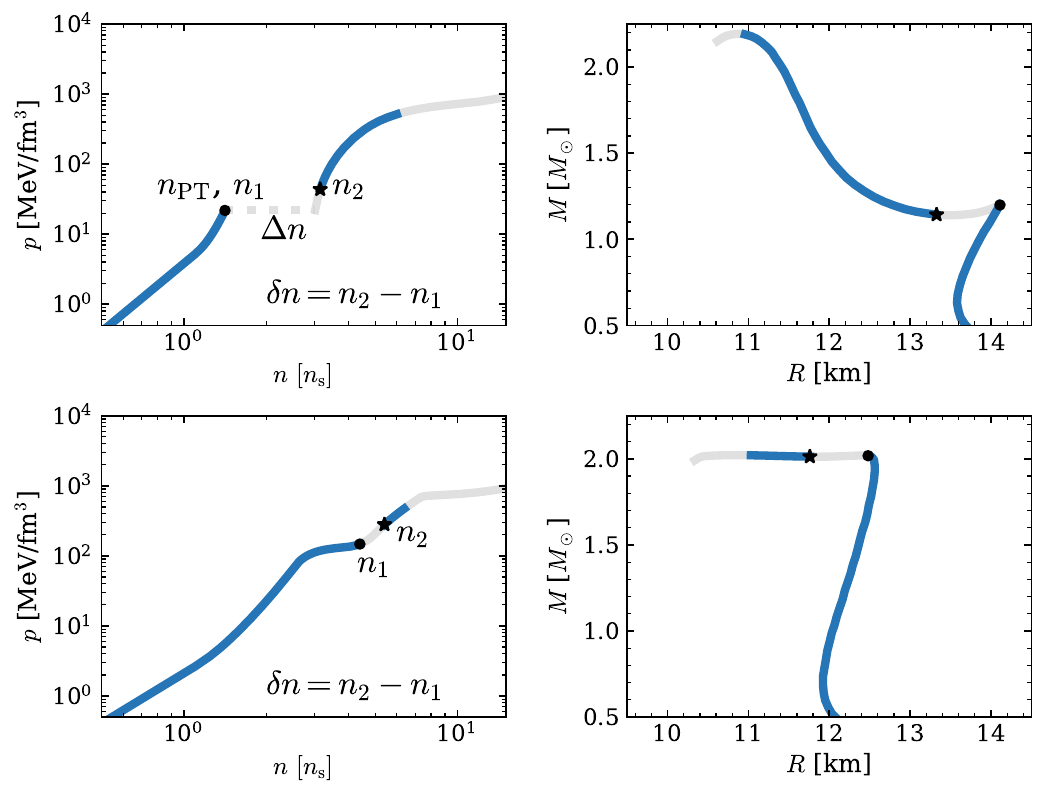}
    \caption{Examples of the two viable twin-star classes 1 (upper row) and 2 (bottom row), including both the EoSs (left column) and corresponding mass–radius curves (right column). Stable stellar branches are shown in blue, unstable ones in gray, and the transition region in the EoS plot in dashed gray. Black dots and stars mark the end of the first and the start of the second stable branch, respectively.
    }
    \label{fig:deltavsdelta}
\end{figure}

In this first Appendix, let us discuss the relation between the phase transition parameters $n_\mathrm{PT}$ and $\Delta n$ as well as the quantities $n_1$, $\delta n$, and $n_2\coloneq n_1+\delta n$.
The physical meaning of these parameters can be illustrated using our two classes of viable twin-star solutions, of which two examples are displayed together with the associated mass-radius curves in \cref{fig:deltavsdelta}.

In the upper left corner of the figure, we show a typical class-1 EoS that features an explicit FOPT at a low density $n_\mathrm{PT}\approx 1.5~n_\mathrm{s}$.
The transition immediately destabilizes the first branch of compact stars, so that the maximal baryon density $n_1$ satisfies the relation $n_1=n_\mathrm{PT}$; we label this point by a black dot in both the EoS and mass–radius plots.
The pressure continues to increase at a somewhat higher density $n_\mathrm{PT}+\Delta n$, which, however, is typically \emph{not} the density, where the second branch begins.
Instead, stability is regained only at an even higher density $n_2=n_\mathrm{PT}+\delta n > n_\mathrm{PT}+\Delta n$, where the pressure becomes sufficient to counteract gravitational collapse.
This marks the onset of the second stable branch, indicated by a black star symbol in both the EoS and mass-radius plots.

The second class of twin stars is displayed on the bottom row of Fig.~\ref{fig:deltavsdelta}. 
It is characterized by an EoS featuring a broad crossover transition centered here around $3-4\,  n_{\rm s}$, where the pressure increases only slowly with density.
This leads to the destabilization of the first branch at $n = n_1\approx 4.5\, n_s$ (black dot), corresponding to stars close to the two-solar-mass bound.
The second branch emerges beyond the crossover at $n = n_2$ (black star), where the pressure has again increased enough to permit stable configurations.
As can be seen from Fig.~\ref{fig:constraints}, many EoSs in the second class exhibit a FOPT at even higher densities, which is typically necessary to reach the pQCD limit without breaking causality and thermodynamic consistency.

\section{Observations \label{sec:obs}}

Next, we briefly review the observations used in the Bayesian calculation described in the main text. The data entering \cref{eq:posterior} include NS mass measurements from radio observations \cite{NANOGrav:2017wvv,Fonseca:2016tux,Demorest:2010bx}, the tidal-deformability parameter $\tilde{\Lambda}$ inferred from the GW170817 binary-neutron-star merger \cite{LIGOScientific:2018hze}, and combined mass-radius inferences from various X-ray observations (including NICER, qLMXB, and X-ray-burster data). 
We summarize the data used in \cref{table:measurements} and refer the interested reader to \cite{Annala:2023cwx} for a detailed summary of the related methodology.

\begin{table}[b]
\begin{tabular}{c c}
\toprule
    System & Refs. \\
    \midrule 
\multicolumn{2}{c}{Radio measurements \vspace{1mm}} \\
   PSR J01614$-$2230 & \cite{NANOGrav:2017wvv,Fonseca:2016tux,Demorest:2010bx} \\
    \midrule 
\multicolumn{2}{c}{NICER pulsars\vspace{1mm}} \\
   PSR J0030+0451 & \cite{Miller:2019cac,Riley:2019yda} \\
   PSR J0740+6620 & \cite{Fonseca:2021wxt,Miller:2021qha,Riley:2021pdl} \\
   PSR J0437$-$4715 & \cite{Choudhury:2024xbk}\\
   PSR J0614$-$3329 & \cite{Mauviard:2025dmd}\\
\midrule
\multicolumn{2}{c}{qLMXB systems\vspace{1mm}} \\
    M13 & \cite{Shaw:2018wxh} \\
    M28, M30,  & \cite{Steiner:2017vmg} \\
    $\omega$~Cen & \cite{Steiner:2017vmg} \\
    NGC 6304, 6397 & \cite{Steiner:2017vmg} \\
    47 Tuc X7 & \cite{Steiner:2017vmg} \\
\midrule
\multicolumn{2}{c}{X-ray bursters\vspace{1mm}} \\
    4U 1702$-$429 & \cite{Nattila:2017wtj} \\
    4U 1724$-$307 & \cite{Nattila:2015jra} \\
    SAX J1810.8$-$260 & \cite{Nattila:2015jra} \\  
    \bottomrule
\end{tabular}
\caption{A summary of the mass and simultaneous mass--radius measurements used to condition the EoSs in this work, with qLMXB standing for quiescent low-mass X-ray binaries.
\label{table:measurements}
}
\end{table}

\bibliographystyle{apsrev4-2}
\bibliography{refs.bib}

\end{document}